\documentclass[12pt,aps,prd,showpacs,amsmath,amssymb]{revtex4}
\usepackage{graphicx}
\input epsf
\allowdisplaybreaks
\begin{document}
\title{$B^{\ast}_{s1}B^{\ast}K$ form factor from QCD sum rules}
\author{Chun-Yu Cui, Yong-Lu Liu and Ming-Qiu Huang}
\affiliation{Department of Physics, National University of Defense
Technology, Hunan 410073, China}
\begin{abstract}
In this article, we calculate the form factors and the coupling constant of the $g_{B^{\ast}_{s1}B^{\ast} K}$ vertex in the framework of the three-point QCD sum rules. Three point correlation functions responsible for the vertex are evaluated by considering both $B^{\ast}$ and $K$ mesons as off-shell states. The form factors obtained are different if the $B^{\ast}$ or the $K$ meson is off-shell but give the same coupling constant.
\end{abstract}
\pacs{ 11.55.Hx,  13.75.Lb, 13.25.Ft,  13.25.Hw}

\maketitle

\section{Introduction}
During the last decade, both theoretical and experimental studies on heavy
mesons have received considerable attention. Studying the properties of bottom-strange orbitally excited states can provide tests for various models of quark bound states\cite{Swa06}. Recently, two orbitally excited narrow $B_{s}$ mesons are reported by the CDF collaboration with masses $m_{B_{s1}}=5829.4 \pm 0.7\,\mbox{MeV}$ and $m_{B^{*}_{s2}}=5839.6 \pm 0.7\,\mbox{MeV}$ \cite{CDF}, which are assigned as the $J^P=(1^+,2^+)$ doublet
states in the heavy quark effective theory \cite{Neubert94}. The D0 collaboration confirmed the $B^{*}_{s2}$ state with mass $m_{B^{*}_{s2}}=5839.6 \pm 1.1 (\text{stat.}) \pm 0.7 (\text{syst.})\,\mbox{MeV}$ in fully reconstructed decays to $B^+K^-$ \cite{D0}. In contrast, the  $B_s$ states with spin-parity $J^P=(0^+,1^+)$ still lack experimental evidence because they are supposed to be broad decaying through an $S$-wave transition.

Based on the potential quark models, heavy quark effective theory, and lattice QCD \cite{BBmeson1,BBmeson2,BBmeson4,BBmeson5,BBmeson6,Matsuki05}, masses of the $B_s$ mesons with $(0^+,1^+)$ have been estimated. With QCD sum rules \cite{Wang0712}, Wang studies masses of the bottomed $(0^+,1^+)$ mesons. The calculation indicates that the central values are below the corresponding $BK$ and $B^{\ast}K$ thresholds respectively, which is just like the case of their charmed cousins $D_{s0}$ and $D^{\ast}_{s1}$. This prediction is compatible with speculation in Refs. \cite{BBmeson2,Colang07}. For this reason, the strong decay of $B^{\ast}_{s1}\rightarrow B^*K$ is not predicted to occur. As a consequence, the width of $B^{\ast}_{s1}$ is very narrow since the main decay channel for this state is isospin symmetry violating $B^{\ast}_{s1}\rightarrow B_{s}^{\ast}\eta\rightarrow B_{s}^{\ast}\pi^0$. In Ref. \cite{Zhu06}, the isospin-violating strong decay width of $B^{\ast}_{s1}$ is calculated, which is around several tens of keV.

Knowledge of the heavy-heavy-light mesons vertices is of crucial importance to estimate the strength of the hadron interactions. They are fundamental objects of low energy QCD. On the experimental side, the excited $B_{s}$ mesons will be copiously accumulated at the CERN LHC, which makes investigations of their decay characters promising. Their coupling constants can bring important constraints in constructing the meson-meson potentials and help us to analyze results of existing experiments in the framework of the meson exchange model~\cite{Liu0911}. For example, Faessler $et. al$ used a phenomenological Lagrangian approach to calculate the strong and radiative decay of the $B_s$ doublets with $j^P=(0^+,1^+)$, which were considered as bound states of the $ B K $ and $ B^* K$  mesons respectively~\cite{FGLM2008}. In the process of their calculation, the coupling constant of $B^{\ast}_{s1}B^{\ast}K$ vertex was an necessary input parameter. However, such low-energy hadron interaction lie in a region which is very far away from the perturbative regime, precluding us to use the perturbative approach with the fundamental QCD Lagrangian. Therefore, we need some non-perturbative approaches, such as QCD sum rules \cite{Shifman,RRY}, to deal with processes relevant to hadron physics. Taking the point of view that the bottomed $(0^+,1^+)$ mesons $B_{s0}$ and $B^{\ast}_{s1}$ are the conventional $b\bar{s}$ mesons, strong coupling constants $g_{B_{s0}BK}$ and $g_{B^{\ast}_{s1}B^{\ast}K}$ are calculated with the light-cone QCD sum rules \cite{Wang2}. The bottomed $0^+$ meson $B_{s0}$ is identified with the conventional $b\bar{s}$ meson and $B_{s0}B K $ vertex has been calculated in Ref. \cite{Nielsen}.

In this article, the form factor and the coupling constant of the $g_{B^{\ast}_{s1}B^{\ast} K}$ vertex is calculated in the framework of the three-point QCD sum rules. In the calculations
we use the same technique developed in the
previous works for the evaluation of the coupling constants in the vetices
$D^* D \pi$ \cite{nnbcs00,nnb02},
$D D \rho$ \cite{bclnn01}, $D^* D^* \pi$ \cite{cdnn05}, $D_s D^* K$, $D_s^* D K$ \cite{bccln06},
$D D \omega$  \cite{hmm07}, $D^*D^*\rho$ \cite{bcnn08}, $D_{sj} D K $
\cite{ko} and $D_s D K^{\ast}_{0}$~\cite{Azizi11}.

This paper is organized as follows. In Sec. \ref{sec2}, we give the details of the QCD sum rules for the considered vertex when both $B^{\ast}$ and $K$ mesons are off-shell. Sec. \ref{sec3} is devoted to the numerical analysis and we compare them with results obtained in other works.

\section{The sum rule for the $B^{\ast}_{s1}B^{\ast}K$ vertex}\label{sec2}

In this section, we obtain QCD sum rules for the form factor of the $B^{\ast}_{s1}B^{\ast} K$ vertex.
The three-point function associated with the $B^{\ast}_{s1}B^{\ast} K$ vertex,
for an off-shell $B^{\ast}$ meson, is given by
\begin{equation}
\Gamma_{\mu\nu\alpha}^{B^{\ast}}(p,p^{\prime})=\int d^4x \, d^4y \;\;
e^{ip^{\prime}\cdot x} \, e^{-iq\cdot y}
\langle 0|T\{j_{\alpha}^{K}(x) j_{\nu}^{B^{\ast}}(y)
 j_{\mu}^{B^{\ast}_{s1}\dagger}(0)|0\rangle,
\label{correboff}
\end{equation}
where the interpolating currents are $j_{\alpha}^{K}(x) = \bar s(x) \gamma_{\alpha} \gamma_{5} u(x)$, $j_{\nu}^{B^{\ast}}(x)= \bar u(x) \gamma_{\nu}b(x)$, and $j_{\mu}^{B^{\ast}_{s1}}(x) = \bar s(x)\gamma_{\mu} \gamma_{5} b(x)$.
The correlation function for an off-shell ${K}$ meson is
\begin{equation}
\Gamma_{\mu\nu}^{{K}}(p,p^{\prime})=\int d^4x \,
d^4y \;\; e^{ip^{\prime}\cdot x} \, e^{-iq\cdot y}\;
\langle 0|T\{j_{\nu}^{B^{\ast}}(x)  j^{K}(y)
 j_{\mu}^{B^{\ast}_{s1} \dagger}(0)\}|0\rangle\, ,\label{correkoff}
\end{equation}
with the interpolating currents $j_{\nu}^{B^{\ast}}(x)= \bar u(x) \gamma_{\nu}b(x)$, $j^{K}(x) = \bar s(x)\gamma_{5} u(x)$, and $j_{\mu}^{B^{\ast}_{s1}}(x) = \bar s(x)\gamma_{\mu} \gamma_{5} b(x)$. In the expressions, $q=p'-p$ is the transferred momentum.

The general expression for Eq. (\ref{correboff})
has fourteen independent Lorentz structures, which can be decomposed in terms of the invariant amplitudes associated
with each of these tensor structures in the following form:
\begin{eqnarray}
\Gamma_{\mu\nu\alpha}^{B^{\ast}}(p,p^{\prime})&=&
    \Gamma_1(p^2 , {p^{\prime}}^2 , q^2) g_{\mu \nu} p_{\alpha}
  + \Gamma_2(p^2,{p^{\prime}}^2, q^2) g_{\mu \alpha} p_{\nu}
  + \Gamma_3(p^2,{p^{\prime}}^2 , q^2) g_{\nu \alpha} p_{\mu}\nonumber \\
&&
  + \Gamma_4(p^2,{p^{\prime}}^2 ,q^2) g_{\mu \nu} p^{\prime}_{\alpha} + \Gamma_5(p^2, {p^{\prime}}^2 ,q^2) g_{\mu \alpha} p^{\prime}_{\nu}
  + \Gamma_6(p^2,{p^{\prime}}^2 ,q^2) g_{\nu\alpha} p^{\prime}_{\mu}\nonumber \\
&&
  + \Gamma_7(p^2,{p^{\prime}}^2 ,q^2) p_{\mu} p_{\nu} p_{\alpha}
  + \Gamma_8(p^2,{p^{\prime}}^2 ,q^2) p^{\prime}_{\mu} p_{\nu} p_{\alpha} + \Gamma_9(p^2,{p^{\prime}}^2 ,q^2) p_{\mu} p^{\prime}_{\nu} p_{\alpha}\nonumber \\
&&
  + \Gamma_{10}(p^2,{p^{\prime}}^2 ,q^2) p_{\mu} p_{\nu} p^{\prime}_{\alpha}
  + \Gamma_{11}(p^2,{p^{\prime}}^2 ,q^2) p^{\prime}_{\mu} p^{\prime}_{\nu} p_{\alpha}
  + \Gamma_{12}(p^2,{p^{\prime}}^2 ,q^2) p^{\prime}_{\mu} p_{\nu} p^{\prime}_{\alpha}  \nonumber \\
&&+ \Gamma_{13}(p^2,{p^{\prime}}^2 ,q^2) p_{\mu} p^{\prime}_{\nu}p^{\prime}_{\alpha}
  + \Gamma_{14}(p^2,{p^{\prime}}^2 ,q^2) p^{\prime}_{\mu} p^{\prime}_{\nu} p^{\prime}_{\alpha}.
\label{trace}
\end{eqnarray}
In principle, we can work with any Dirac structure appearing  in Eqs.~(\ref{trace}). However, there are still some points
which one must follow: (i) The chosen structure must appear in the phenomenological side. (ii) The chosen structure should exhibit good OPE convergence. (iii) The chosen structure should have
a stability that guarantees a good match between the two sides of the sum rule. After some calculations, we find that the structures which obey these  points are $p^{\prime}_{\mu}p^{\prime}_{\nu} p^{\prime}_{\alpha}$ in the case of $B^{\ast}$ off-shell, and $p^{\prime}_{\mu}p^{\prime}_{\nu}$ in
the case of K off-shell.

In order to get the sum rules, the correlation functions need to be calculated in two different ways: in phenomenological or physical side, they are presented in terms of hadronic parameters such as masses, leptonic
decay constants and the form factor; in theoretical or QCD side,
they are evaluated in terms of QCD degrees of freedom via operator product expansion (OPE) in deep Euclidean region. The sum rules for the form factors are obtained with both representations being matched, invoking
the quark-hadron global duality and equating the coefficient of a sufficient structure from both sides of the same correlation functions.
To improve the matching between the two representations, double Borel transformation with respect to the variables, $P^2=-p^2\rightarrow M^2$ and ${P^\prime}^2=-{p^\prime}^2\rightarrow {M^{\prime}}^2$, is performed.

In detail, we calculate the physical part of the first
correlation function (\ref{correboff}) for an off-shell $B^{\ast}$ meson. The form factor $g_{B^{\ast}_{s1}B^{\ast}K}(q^2)$ is defined by the matrix element
\begin{equation}
\langle B^{\ast}_{s1}(p)|K(p') B^{\ast}(q)\rangle =-i g_{B^{\ast}_{s1}B^{\ast}K}(q^2)\eta^* \cdot
\epsilon,
\label{ffb}
\end{equation}
where $\epsilon$ and $\eta$ are the polarization
vectors associated with the $B^{\ast}$ and $B^{\ast}_{s1}$,
respectively.
The meson decay constants $f_{B^{\ast}_{s1}}$, $f_{B^{\ast}}$, and $f_{K}$ are
defined by the following matrix elements:
\begin{eqnarray}
\langle 0|j_{\mu}^{B^{\ast}_{s1}}|{B^{\ast}_{s1}(p)}\rangle&=&m_{B^{\ast}_{s1}} f_{B^{\ast}_{s1}}\eta_{\mu},\nonumber\\
\langle 0|j_{\nu}^{B^{\ast}}|{B^{\ast}(p)}\rangle&=&m_{B^{\ast}} f_{B^{\ast}}\epsilon_{\nu},\nonumber\\
\langle 0|j_{\alpha}^{K}|{K(p)}\rangle&=&i f_K p_{\alpha}.
\label{fK2}
\end{eqnarray}
Saturating Eq.~(\ref{correboff}) with
$B^{\ast}_{s1}$, $B^{\ast}$ and $K$ states, using Eqs.~(\ref{ffb}) and (\ref{fK2}), and then summing over polarization vectors via
\begin{eqnarray}\label{polvec}
\epsilon_\mu\epsilon^*_\nu=-g_{\mu\nu}+\frac{q_{\mu}q_{\nu}}{m_{B^{\ast}}^2},\nonumber\\
\eta_\mu\eta^{*}_\nu=-g_{\mu\nu}+\frac{p_{\mu}p_{\nu}}{m_{B^{\ast}_{s1}}^2},
\end{eqnarray}
the physical
side of the correlation function for $B^{\ast}$  off-shell can be
obtained as
\begin{eqnarray}
\Gamma_{\mu\nu\alpha }^{B^{\ast}phen}(p,p^{\prime})&=& g^{B^{\ast}}_{B^{\ast}_{s1} B^{\ast} K}(q^2)
 \frac{f_{B^{\ast}_{s1}} f_{K} f_{B^{\ast}} m_{B^{\ast}}m_{B^{\ast}_{s1}}}
{(p^2-m^2_{B^{\ast}_{s1}})(q^2-m^2_{B^{\ast}})({p^{\prime}}^2 -m^2_{K})} \times\nonumber\\
&&\left(-g_{\mu\gamma}+\frac{p_{\mu}p_{\gamma}} {m^2_{B^{\ast}_{s1}}}\right)\left(-g_{\nu\gamma}+\frac{q_{\nu}q_{\gamma}} {m^2_{B^{\ast}}}\right)p^{\prime}_{\alpha}+...\,.
\label{phenboff}
\end{eqnarray}

A similar procedure is carried out to obtain the final expression of the physical side of the correlation function for an off-shell $K$ meson as
\begin{eqnarray}
\Gamma_{\mu\nu}^{K phen}(p,p^{\prime})&=& g^{K}_{B^{\ast}_{s1} B^{\ast} K}(q^2)
 \frac{f_{B^{\ast}_{s1}} f_{K} f_{B^{\ast}} \frac{m^2_{K}}{m_s} m_{B^{\ast}_{s1}} m_{B^{\ast}}}
{(p^2-m^2_{B^{\ast}_{s1}})(q^2-m^2_{K})({p^{\prime}}^2 -m^2_{B^{\ast}})} \times\nonumber\\
&&\left(-g_{\mu\gamma}+\frac{p_{\mu}p_{\gamma}} {m^2_{B^{\ast}_{s1}}}\right)\left(-g_{\nu\gamma}+\frac{p^{\prime}_{\nu}p^{\prime}_{\gamma}} {m^2_{B^{\ast}}}\right)+...\,,
\label{phenkoff}
\end{eqnarray}
in which ``..." represents the contributions of the higher states and
continuum.

\begin{figure}
  \includegraphics[width=15cm]{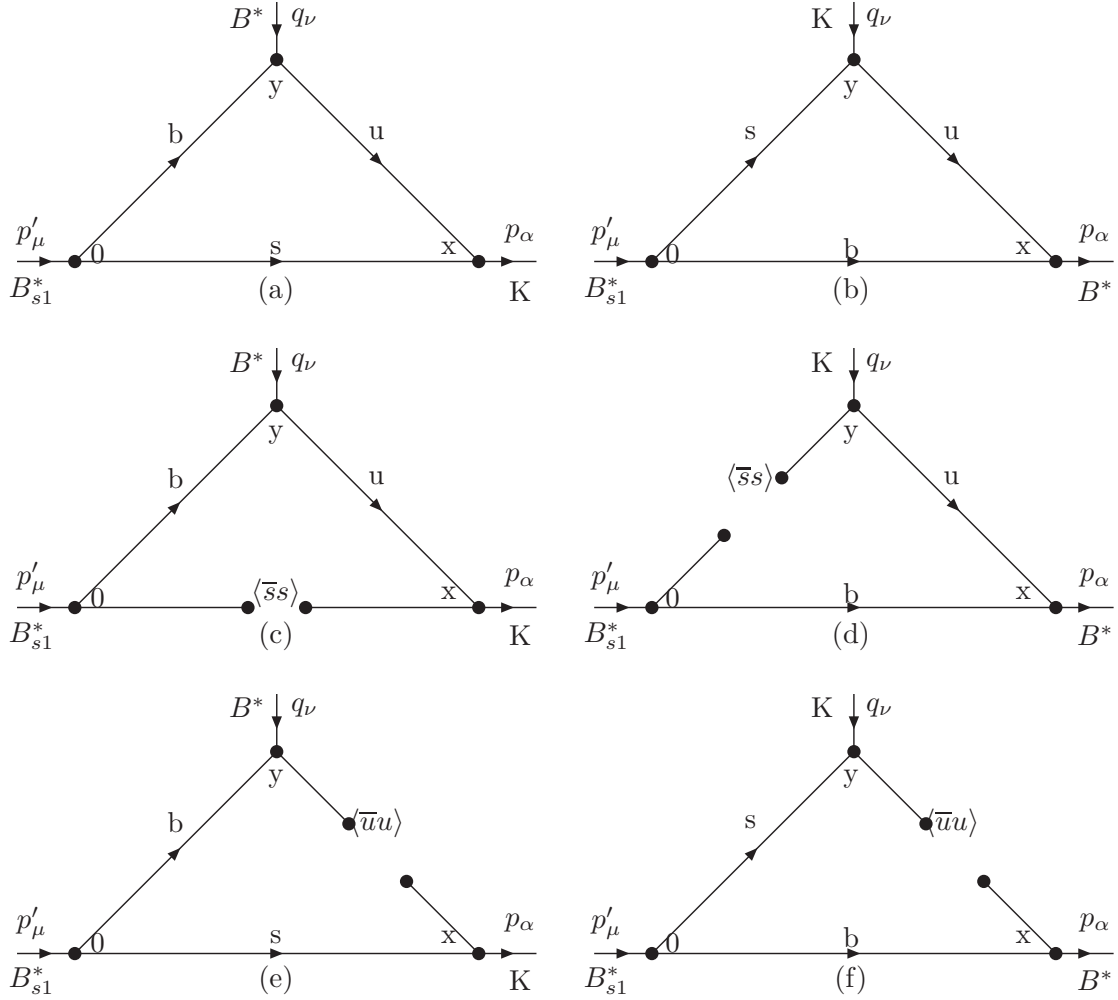}\\
  \caption{\quad {(a) and (b): Bare loop diagram for the $B^{\ast}$  and
$K$ off-shell, respectively; (c) and (e): Diagrams
corresponding to quark condensate for the $B^{\ast}$ off-shell; (d) and
(f): Diagrams corresponding to quark condensate for the  $K$
off-shell.}}
\label{Figure1}
\end{figure}

In the following, we turn to the QCD side of the correlation
functions in deep Euclidean space. For an off-shell $B^{\ast}$ meson, each invariant amplitude $\Gamma^{i}(p^{\prime},p)$ appearing in Eq.~(\ref{trace}) can be written in terms of
perturbative and non-perturbative parts:
\begin{eqnarray}\label{CorrelationFuncQCD}
\Gamma^{i}(p^{\prime},p)&=&
\left(\Pi^{i}_{per}+\Pi^{i}_{nonper}\right).
\end{eqnarray}
In the expression, the perturbative part is defined as
\begin{equation}
\Pi^{i}_{per}=-\frac{1}{4\pi^2}\int_{s_{min}}^\infty ds
\int_{u_{min}}^\infty du \:\frac{\rho_i(s,u,Q^2)}{(s-p^2)(u-{p^\prime}^2)}\;,
\;\;\;\;\;\;i=1,\ldots,14, \label{dis}
\end{equation}
where $\rho_i(s,u,Q^2)$ is called spectral density. The spectral density is obtained by calculating the bare
loop diagrams (a) and (b) in Fig. (\ref{Figure1}) for $B^{\ast}$ and
$K$ off-shell, respectively. We calculate these
diagrams in terms of the usual Feynman integral with
Cutkosky rules, i.e., by replacing the quark propagators with Dirac
delta function $\frac{1}{q^2-m^2}\rightarrow (- 2\pi i)
\delta(q^2-m^2)\theta(q^{0})$. After some straightforward calculations, the spectral density associated with the structure $p^{\prime}_{\mu}p^{\prime}_{\nu} p^{\prime}_{\alpha}$ is obtained as following:
\begin{align}
\rho^{B^{\ast}}(s,t,u)=&\frac{12}{[\lambda(s,u,t)]^{7/2}}[m_{b}^6 (s^3-3 s^2 (t-3 u)+3 s (t^2-4 t u+3 u^2)-(t-u)^3)\nonumber \\
&+m_{b}^4 (3 m_{s}^2 (3 s^3+s^2 (3 u-5 t)+s (t^2+4 t u-5 u^2)+(t-u)^3)-2 s^4+s^3 (5 t-13 u)\nonumber \\
&+s^2 (-3 t^2+4 t u+3 u^2)-s (t-11 u) (t-u)^2+(t-u)^4)\nonumber \\
&+m_{b}^2 (3 m_{s}^4 (3 s^3+s^2 (3 t-5 u)+s (-5 t^2+4 t u+u^2)-(t-u)^3)\nonumber \\
&-2 m_{s}^2 (4 s^4-s^3 (t+u)+s^2 (-9 t^2+22 t u-9 u^2)+5 s (t-u)^2 (t+u)+(t-u)^4)\nonumber \\
&+s (s^4-s^3 (t-5 u)+s^2 (-3 t^2+10 t u-9 u^2)+s (5 t^3-17 t^2 u+13 t u^2-u^3)\nonumber \\
&-2 (t-u)^3 (t+2 u)))+m_{s}^6 (s^3+s^2 (9 t-3 u)+3 s (3 t^2-4 t u+u^2)+(t-u)^3)\nonumber \\
&+m_{s}^4 (-2 s^4+s^3 (5 u-13 t)+s^2 (3 t^2+4 t u-3 u^2)+s (t-u)^2 (11 t-u)+(t-u)^4)\nonumber \\
&+m_{s}^2 s (s^4+s^3 (5 t-u)+s^2 (-9 t^2+10 t u-3 u^2)-s (t^3-13 t^2 u+17 t u^2-5 u^3)\nonumber \\
&+2 (t-u)^3 (2 t+u))+s^2 (s^3 (-(t+u))+s^2 (3 t^2-2 t u+3 u^2)\nonumber \\
&+s (-3 t^3+t^2 u+t u^2-3 u^3)+(t-u)^2 (t^2+4 t u+u^2))]
\label{SpecDenOSB}
\end{align}
With a similar procedure, we obtain the spectral density
\begin{align}
\rho^{K}(s,t,u)=&\frac{6m_{s}}{[\lambda(s,u,t)]^{5/2}}[m_{b}^4(s^2-2 u (s+t)+4 s t+t^2+u^2)
\nonumber \\
&+m_{b}^2(m_{s}^2(4s2-2 s (t+u)-2(t-u)^2)-(s-t+u)(s^2-2 u (s+t)\nonumber \\
&+4 s t+t^2+u^2)) +m_{s}^4((s-t)^2+4 s u-2 t u+u^2)\nonumber \\
&-m_{s}^2 (s+t-u) ((s-t)^2+4su-2t
u+u^2)\nonumber \\
&+s(s^2 (t+u)-2 s (t^2-t u+u^2)+(t-u)^2 (t+u))]
\label{SpecDenOSK}
\end{align}
for the off-shell $K$ meson associated with the structure $p^{\prime}_{\mu}p^{\prime}_{\nu}$. All powers of the strange quark mass are included when calculating the spectral density. Herein
$\lambda(a,b,c)=a^2+b^2+c^2-2ac-2bc-2ab$ and $t=q^2=-Q^2$.

The nonperturbative contributions in the QCD side are determined from the quark condensate diagrams (c), (d), (e) and (f) of Fig.(\ref{Figure1}). As what has been shown in Refs.~\cite{nnbcs00,Nielsen}, heavy quark condensate and gluon condensate contributions are negligible as compared with the perturbative one. As a result, only light quark condensates contribute to the calculation. It is noticed that contributions of diagrams (d), (e) and (f) are zero after the double Borel transformation with respect to the both variables $P^2$ and ${P^\prime}^2$ . Hence, we calculate the diagram (c) for the off-shell $B^{\ast} $ meson and obtain
\begin{eqnarray}\label{CorrelationFuncNonpert}
\Pi_{nonper}^{B^{\ast}}&=&
-\frac{m_{b}{\langle}\overline{s}s{\rangle}}{(p^2-m_{b}^2){p^{\prime}}^2}(g_{\mu\nu}p^{\prime}_{\alpha}+g_{\mu\alpha}p^{\prime}_{\nu}-g_{\nu\alpha}p^{\prime}_{\mu})
\end{eqnarray}
for the off-shell $B^{\ast}$ meson.

We use the quark-hadron duality assumption to subtract the contributions of the higher states and continuum, i.e., it is assumed that
\begin{eqnarray}\label{ope}
\rho^{higher states}(s,u) = \rho^{OPE}(s,u,t) \theta(s-s_0)
\theta(u-u_0),
\end{eqnarray}
where $s_0$ and $u_0$ are the continuum thresholds.

The next step of the method is to apply the double Borel transformation with respect
to the $P^2=-p^2\rightarrow M^2$ and ${P^\prime}^2=-{p^\prime}^2\rightarrow {M^{\prime}}^2$ to the physical as well
as the QCD side. We get the final sum rules for the corresponding form factors as
\begin{eqnarray}\label{boff}
g^{B^{\ast}}_{B^{\ast}_{s1}B^{\ast}K}(Q^2)&=&\frac{3m_{B^{\ast}}(Q^2+m_{B^{\ast}}^2)}{f_{B^{\ast}_{s1}}
f_{B^{\ast}} f_{K}m_{B^{\ast}_{s1}}}
e^{\frac{m_{B^{\ast}_{s1}}^2}{M^2}}e^{\frac{m_{K}^2}{{M^{\prime}}^2}}
\left[\frac{1}{4~\pi^2}\int^{s_0}_{(m_{b}+m_{s})^2}
ds\int_0^{u_{max}} du
\rho^{B^{\ast}}(s,u,t) \right.
\nonumber \\
&& \left.
e^{\frac{-s}{M^2}}e^{\frac{-u}
{{M^{\prime}}^2}}\right]
\end{eqnarray}
and
\begin{eqnarray}\label{koff}
g^{K}_{B^{\ast}_{s1}B^{\ast}K}(Q^2)&=&\frac{3m_{s}m_{B^{\ast}}(Q^2+m_{K}^2)}
{f_{B^{\ast}_{s1}}
f_{B^{\ast}} f_{K}{m_{K}^2}m_{B^{\ast}_{s1}}}
e^{\frac{m_{B^{\ast}_{s1}}^2}{M^2}}e^{\frac{m_{B^{\ast}}^2}{{M^{\prime}}^2}}
\left[\frac{-1}{4~\pi^2}\int^{s_0}_{(m_{b}+m_{s})^2}
ds\int^{u_0}_{u_{min}} du
\rho^{K}(s,u,t) \right.
\nonumber \\
&& \left.
e^{\frac{-s}{M^2}}e^{\frac{-u}
{{M^{\prime}}^2}}\right]
\end{eqnarray}
for the off-shell $B^{\ast}$ and $K$ meson associated with the $B^{\ast}_{s1}B^{\ast}K$ vertex, respectively. In the above expressions, we have $u_{min}=m_{b}^2-\frac{m_{b}^2 t}{s-m_b^2}$ and $u_{max}=s+t-m_{b}^2-\frac{s t}{m_{b}^2}$, and the integration regions are determined from the fact that arguments of the
three $\delta$ functions must vanish simultaneously with consideration of the three Heaviside step functions.

It is noticed that in this work we use the relations between the Borel masses $M^2$ and ${M^{\prime}}^2$ with
$\frac{M^2}{{M^{\prime}}^2} = \frac{m^2_{B^{\ast}}-m^2_{b}}{0.64}$
for a  $B^{\ast}$ off-shell  and $\frac{M^2}{M'^2} = \frac{m^2_{B^{\ast}_{s1}}}{m^2_{B^{\ast}}}$ for
a $K$ off-shell.

\section{Numerical analysis}\label{sec3}
\begin{figure}
\begin{minipage}{7cm}
\epsfxsize=7cm \centerline{\epsffile{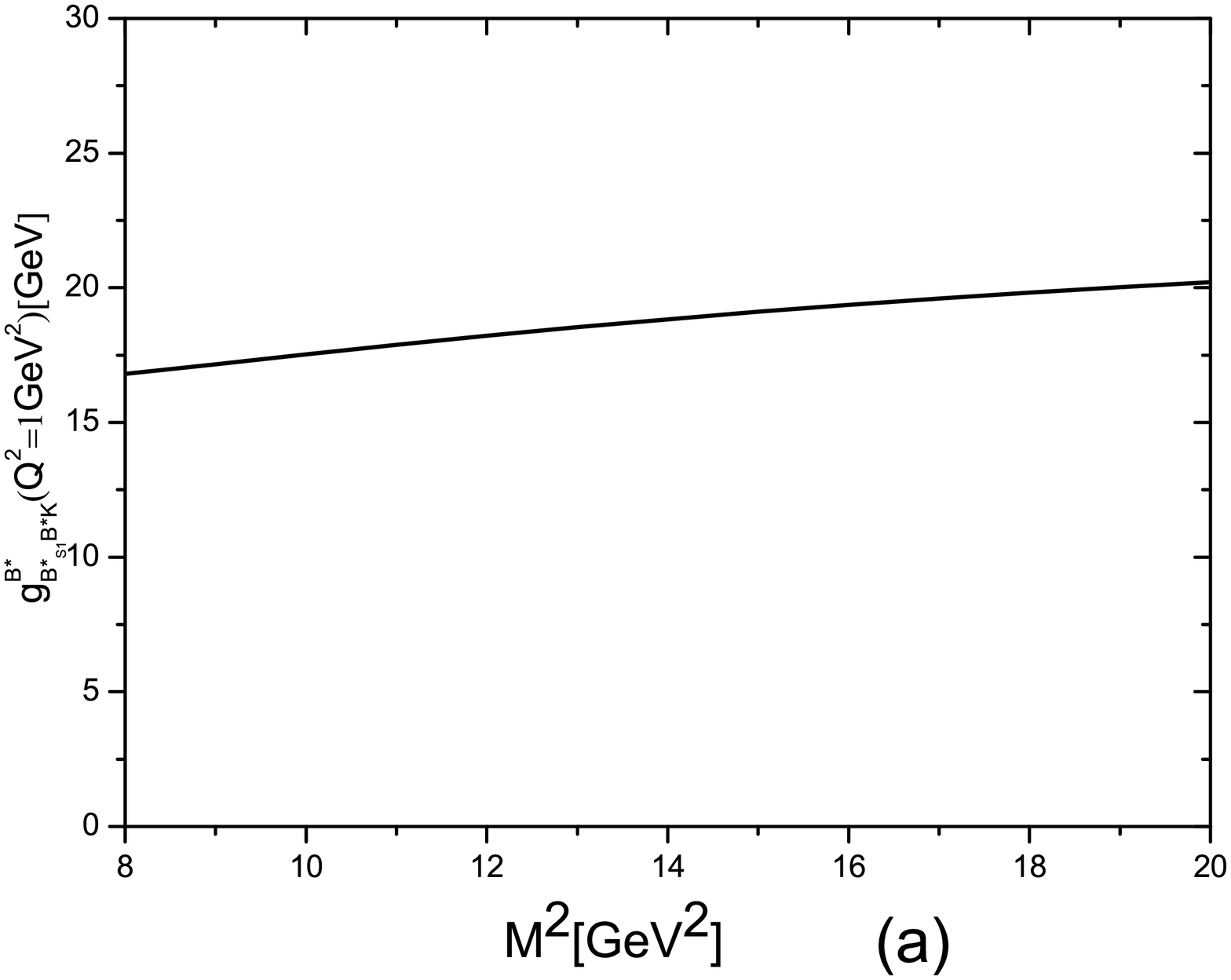}}
\end{minipage}
\begin{minipage}{7cm}
\epsfxsize=7cm \centerline{\epsffile{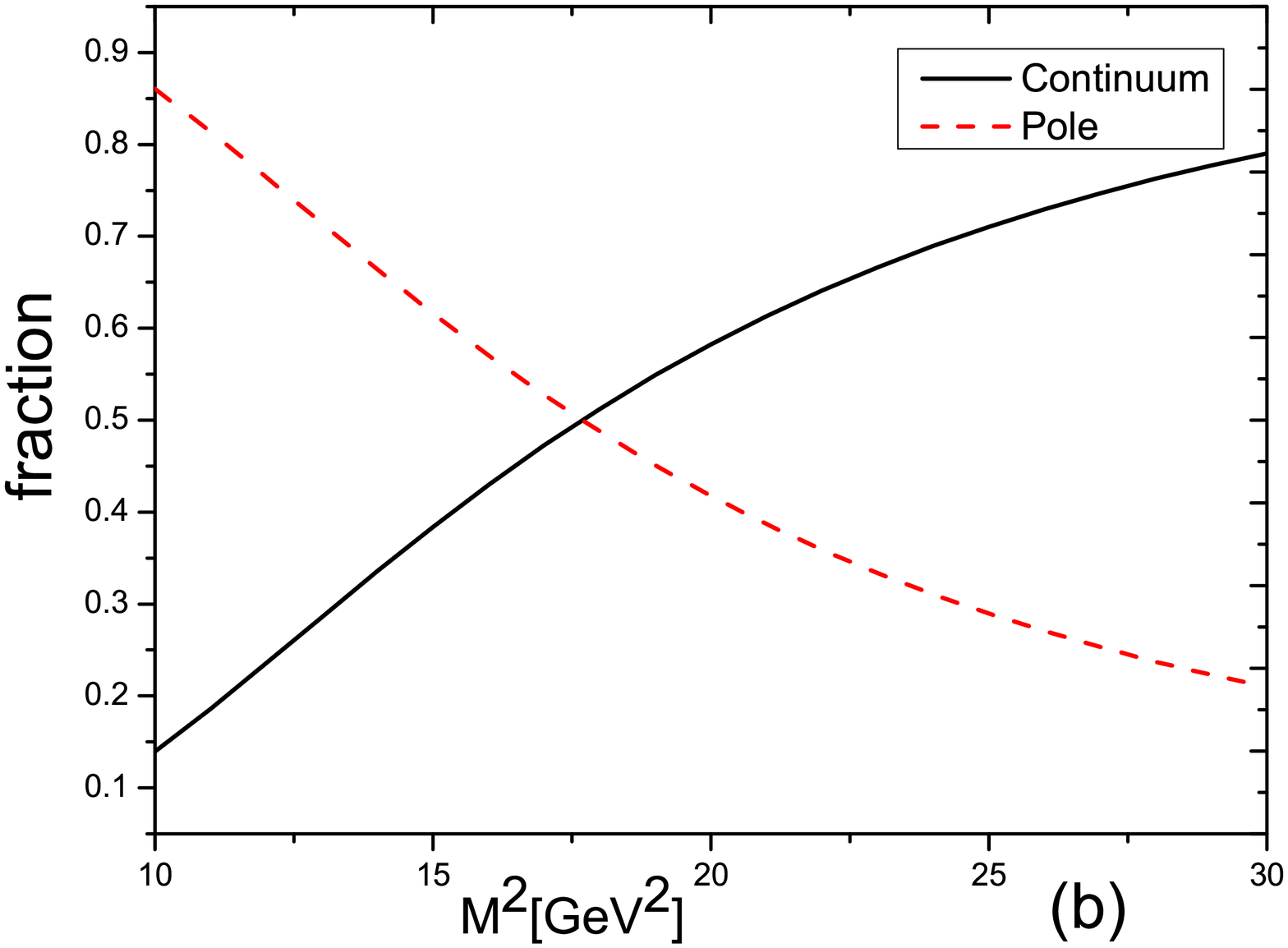}}
\end{minipage}
\caption{a) The dependence of the form factor $g^{B^{\ast}}_{B^{\ast}_{s1}B^{\ast}K}(Q^2=1.0\,\mbox{GeV}^2)$ on Borel mass parameters $M^2$ for $\Delta_{s} =0.5\,\mbox{GeV}$ and $\Delta_{u} =0.5\,\mbox{GeV}$ and b) pole-continuum contributions.}\label{Figure2}
\end{figure}

This section is devoted to the numerical analysis of the sum
rules for the form factor. Input parameters are shown in Table \ref{table1}. The values $f_{B^{\ast}_{s1}}$, $m_{B^{\ast}_{s1}}$ and $m_b$ are used the same as that in Ref.~\cite{Wang0712}. We first determine the three auxiliary parameters in the sum rules, namely the Borel mass parameter $M^2$ and the continuum thresholds, $s_0$ and $u_0$. The continuum thresholds, $s_0$ and $u_0$, are not completely arbitrary as they are correlated to the
energy of the first excited states with the same quantum numbers as the state we concern. They are given by $s_0=(m_{B^{\ast}_{s1}} + \Delta_{s})^2$ and $u_0=(m+\Delta_{u})^2$, where $m$ is the K meson mass for the case that $B^{\ast}$ is off-shell and the $B^{\ast}$ meson mass for that $K$ is off-shell. Since these
parameters are not physical quantities, we need to look for the working regions at which pole dominance and stability of the sum rule with the Borel mass parameter are satisfied. Using $\Delta_{s}=\Delta_{u} = 0.5\,\mbox{GeV} $ for the continuum thresholds
and fixing $Q^2=1\,\mbox{GeV}^2$, we find a good stability of the
sum rule for $g^{B^{\ast}}_{B^{\ast}_{s1}B^{\ast}K}$ with $M^2\geq 10\,\mbox{GeV}^2$, as is shown in
Fig.~(\ref{Figure2}a). Fig.~(\ref{Figure2}b) demonstrates that contributions from pole terms with variation of the Borel parameter $M^2$. We see that the pole contribution is larger than continuum contribution for $M^2\leq 17\,\mbox{GeV}^2$. We choose $M^2= 15\,\mbox{GeV}^2$ as a reference point.

\begin{table}[h]
\caption{\quad Parameters used in the calculation.}
\begin{center}
\begin{tabular}{ccccccccc}
\hline
$m_{s} (\mbox{GeV})$&  $m_{b} (\mbox{GeV})$ & $m_{B^{\ast}_{s1}} (\mbox{GeV})$ & $m_{B^{\ast}} (\mbox{GeV})$ & $m_{K} (\mbox{GeV})$ &  $f_{B^{\ast}_{s1}} (\mbox{GeV})$& $f_{B^{\ast}}(\mbox{GeV})$  & $f_{K} (\mbox{GeV})$ & $\langle \bar s s \rangle(\mbox{GeV})^3$\\
\hline
0.13& 4.7 &5.72 &5.325 &0.49 & 0.24 &0.17 &0.16 &$0.8*(-0.23)^3$ \\ \hline
\end{tabular}
\end{center}\label{table1}
\end{table}

\begin{figure}
\includegraphics[width=15cm]{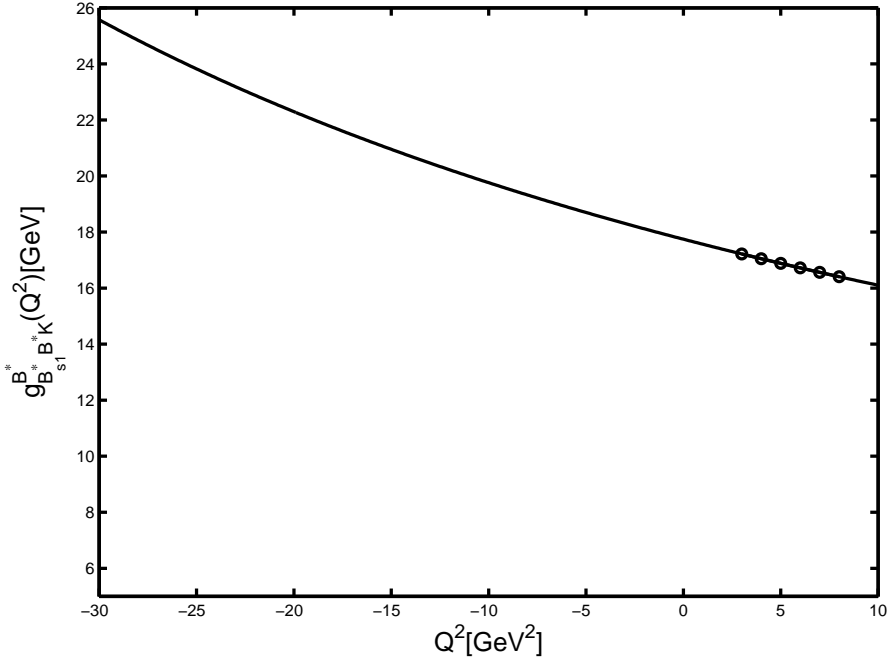}
\caption{\quad $g^{B^{\ast}}_{B^{\ast}_{s1}B^{\ast}K}$ (circles) QCDSR form factors as a function of
$Q^2$. The solid line correspond to the monopolar parametrization of the QCDSR data.}\label{Figure3}
\end{figure}

Now, with the determined $M^2$ and other input parameters, the dependence of the form factors on $Q^2$ is plotted in Fig.~\ref{Figure3}. The circles correspond to the $g^{B^{\ast}}_{B^{\ast}_{s1}B^{\ast}K}(Q^2)$ form factor in the interval where the sum rule is valid. Following previous works \cite{bclnn01,Nielsen}, we fit the circles with $B^{\ast}$ off-shell by the monopolar parametrization,
\begin{equation}
g_{B^{\ast}_{s1}B^{\ast}K}^{B^{\ast}}(Q^2)= \frac{1739\,\mbox{GeV}^3}{Q^2+98\,\mbox{GeV}^2} \;.
\label{gBoff}
\end{equation}
The coupling constant is defined as the value of the form factor at $Q^2=-m^2$, where $m$ is the mass of the off-shell meson.
Using $Q^2=-m_{B^{\ast}}^2$ in Eq. (\ref{gBoff}),
the coupling constant is obtained as $g^{B^{\ast}}_{B^{\ast}_{s1}B^{\ast}K}=24.96~\mbox{GeV}$.

\begin{figure}
\begin{minipage}{7cm}
\epsfxsize=7cm \centerline{\epsffile{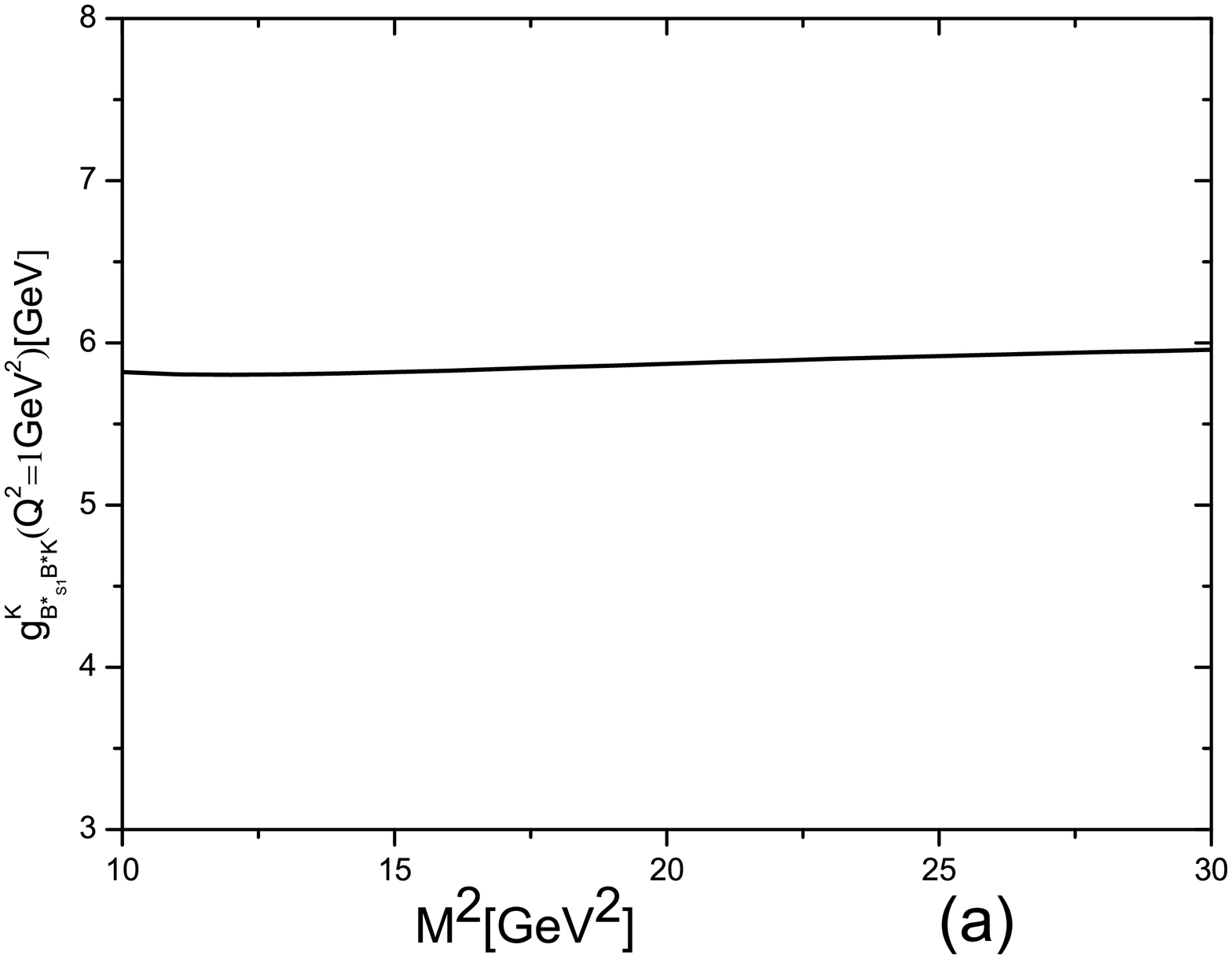}}
\end{minipage}
\begin{minipage}{7cm}
\epsfxsize=7cm \centerline{\epsffile{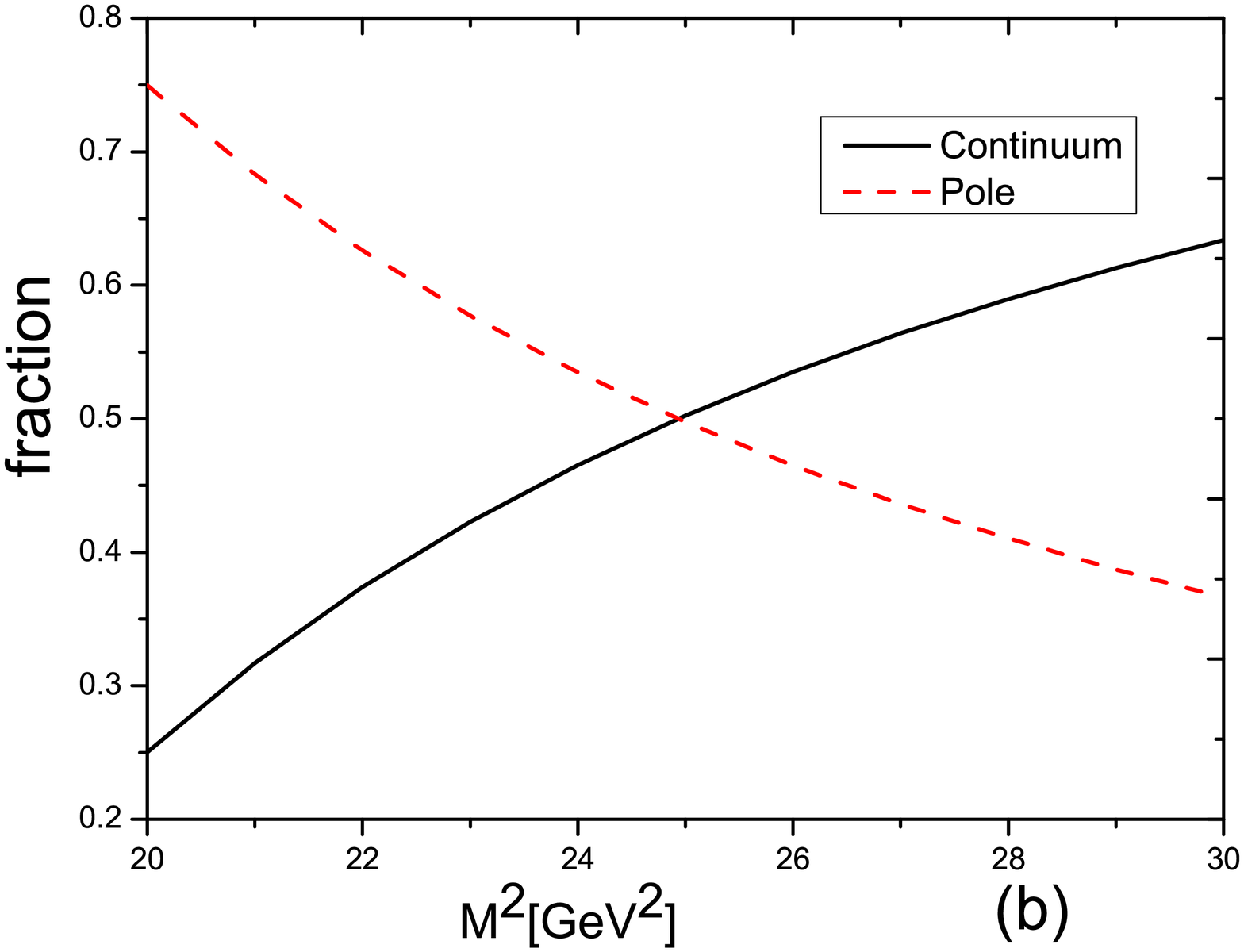}}
\end{minipage}
\caption{a) The dependence of the form factor $g^{K}_{B^{\ast}_{s1}B^{\ast}K}(Q^2=1.0\,\mbox{GeV}^2)$ on Borel mass parameters for $\Delta_{s} =0.5\,\mbox{GeV}$ and $\Delta_{u} =0.5\,\mbox{GeV}$ and b) pole-continuum contributions.}\label{Figure4}
\end{figure}

\begin{figure}
\includegraphics[width=15cm]{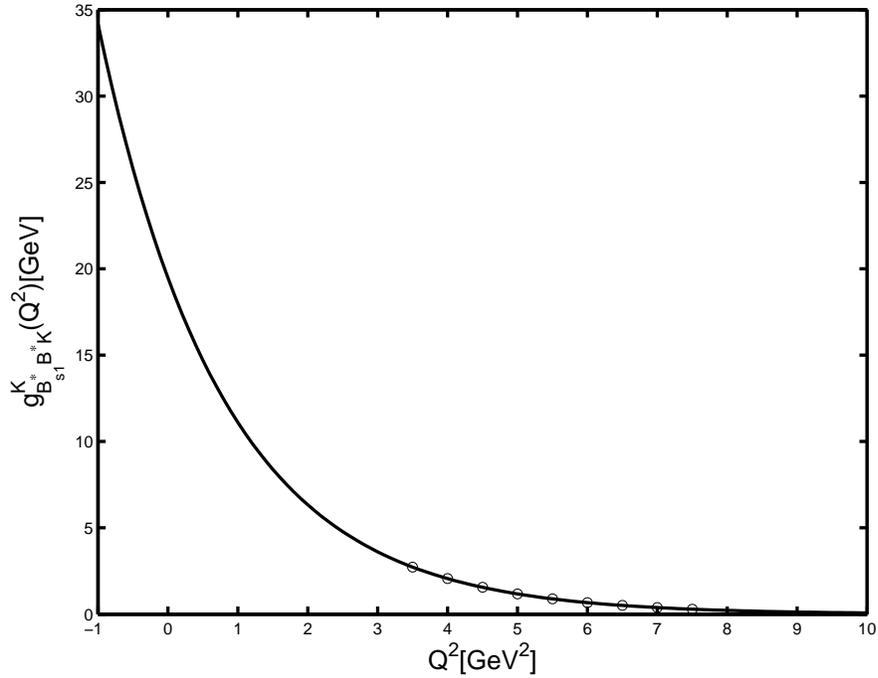}
\caption{\quad $g^{K}_{B^{\ast}_{s1}B^{\ast}K}(Q^2)$ (circles) QCDSR form factors as a function of $Q^2$. The solid line correspond to the exponential parametrization of the QCDSR data.}\label{Figure5}
\end{figure}

In the case that $K$ is off-shell, Fig.~(\ref{Figure4}a) demonstrates a good stability of $g^{K}_{B^{\ast}_{s1}B^{\ast}K}$ with respect to the variations of Borel parameters for $M^2\geq 15\,\mbox{GeV}^2$. We see that the pole contribution is bigger than the continuum one in the Borel window $M^2\leq 25\,\mbox{GeV}^2$ from Fig.~(\ref{Figure4}b). We choose $M^2= 23\,\mbox{GeV}^2$ as a reference point. Our numerical results can be fitted by the exponential parametrization
\begin{eqnarray}\label{grhooff}
g^{K}_{B^{\ast}_{s1}B^{\ast}K}(Q^2)=19.48
\,\mbox{Exp}\Big(\frac{-Q^2}{1.78\,\mbox{GeV}^2}\Big)~\mbox{GeV},
\end{eqnarray}
shown by the solid line in Fig.~\ref{Figure5}. Also, $g^{K}_{B^{\ast}_{s1}B^{\ast}K}=22.29~\mbox{GeV}$ is obtained at $Q^2=-m_{K}^2$ in Eq. (\ref{grhooff}), which is in a good agreement with the result $g^{B^{\ast}}_{B^{\ast}_{s1}B^{\ast}K}=24.96~\mbox{GeV}$. Taking the average of these results, we get
\begin{equation}\label{CoupConstDsDKs}
g_{B^{\ast}_{s1}B^{\ast}K}=(23.62\pm1.33)~\mbox{GeV}.
\end{equation}
Comparing Fig.~\ref{Figure3} with Fig.~\ref{Figure5}, we also discover that the heavier is the off-shell meson, the more stable is its form factor in terms of $Q^2$, which was first observed in Ref.~\cite{bclnn01}.
\begin{figure}
\begin{minipage}{7cm}
\epsfxsize=7cm \centerline{\epsffile{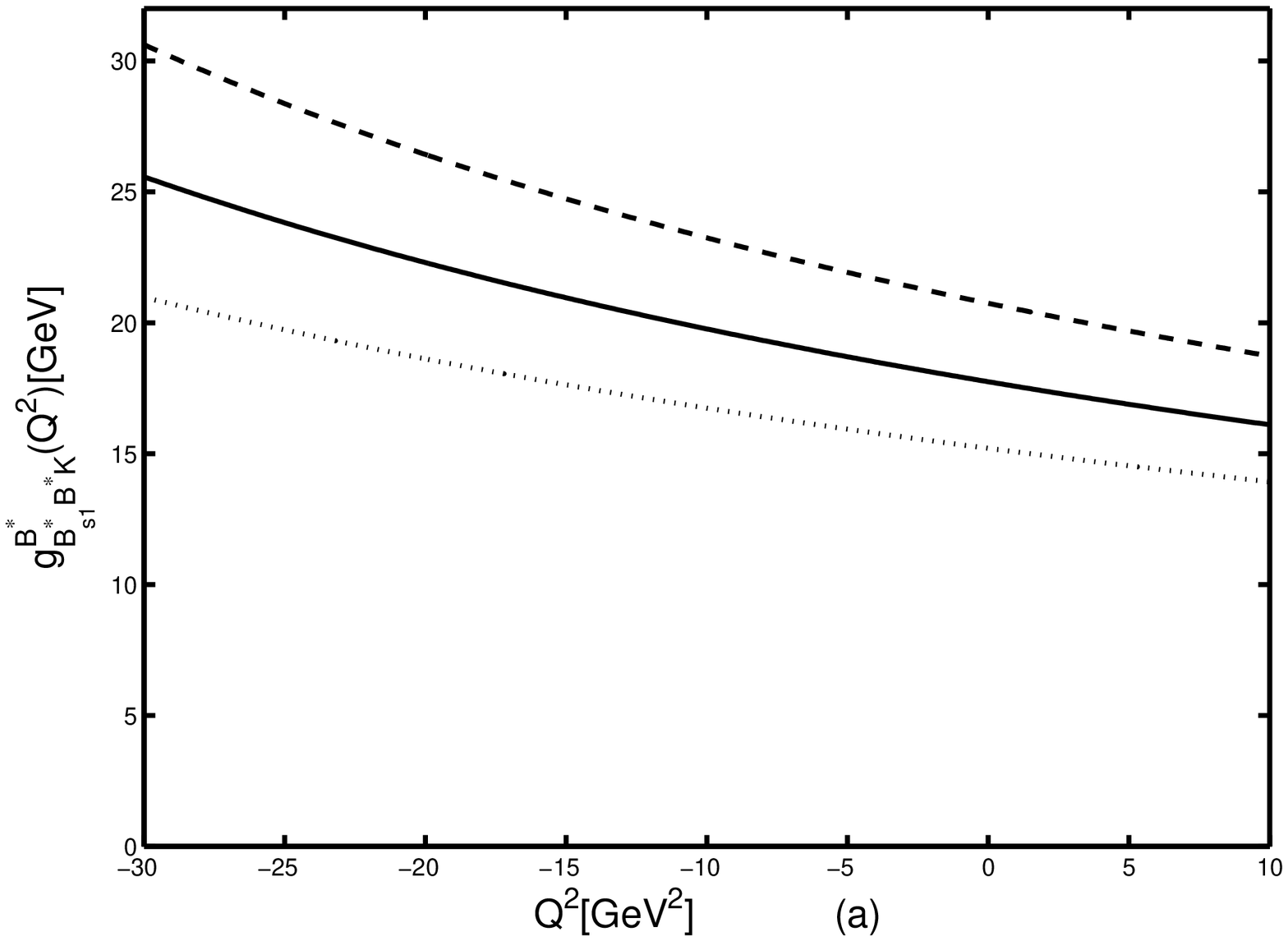}}
\end{minipage}
\begin{minipage}{7cm}
\epsfxsize=7cm \centerline{\epsffile{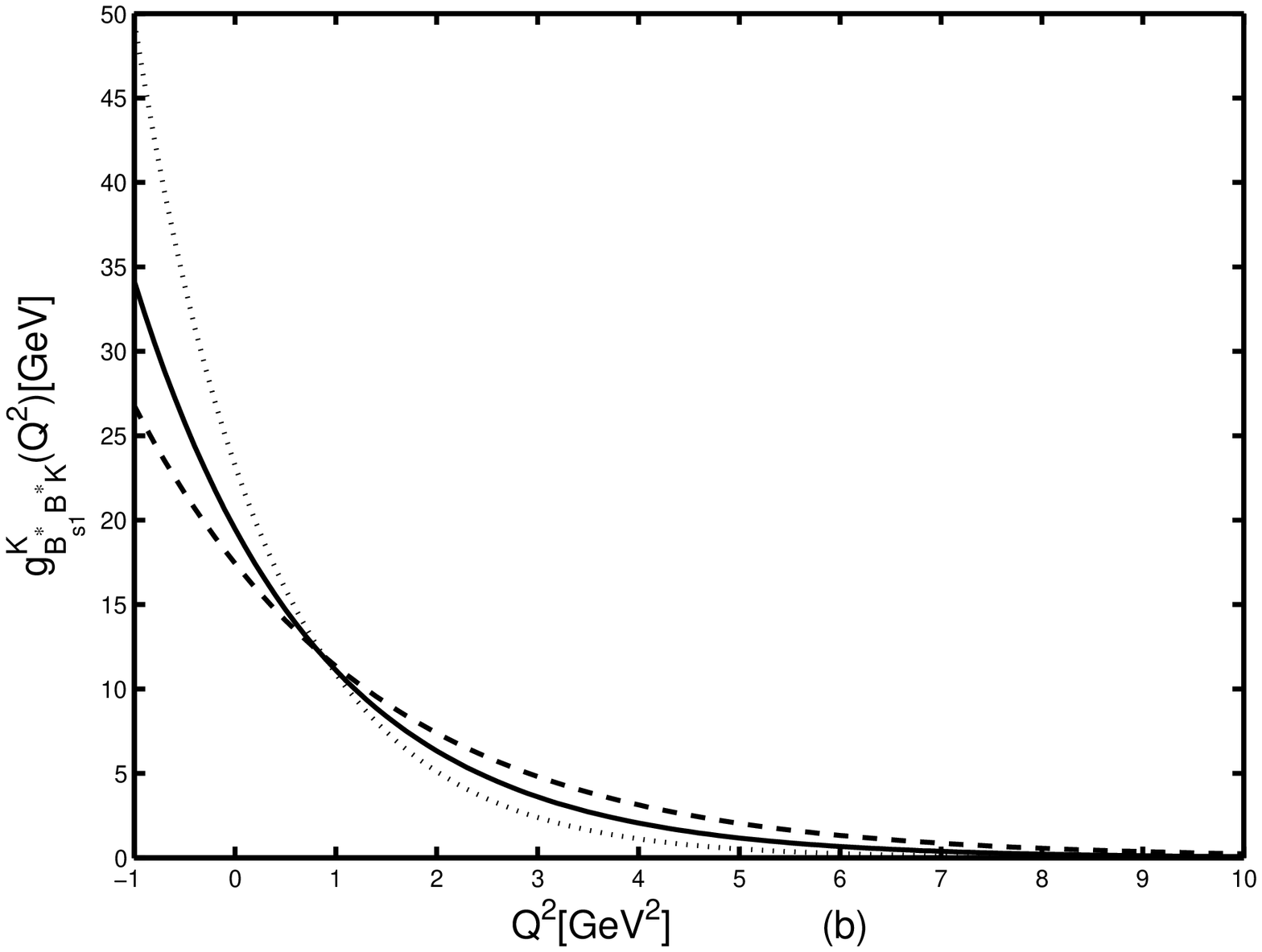}}
\end{minipage}
\caption{\quad Dependence of the form factor $g_{B^{\ast}_{s1}B^{\ast}K}(Q^2)$ on the continuum threshold for
both $B^{\ast}$ and $K$ off-shell cases. The dotted curve corresponds to
$\Delta_{s,u} = 0.4\mbox{GeV}$, the solid one to $\Delta_{s,u}= 0.5\mbox{GeV}$
and the dashed curve to $\Delta_{s,u}= 0.6\,\mbox{GeV}$.}
\label{Figure6}
\end{figure}

In order to study the dependence of our results on the continuum threshold, we vary $\Delta_{s,u}$ between $0.4\,\mbox{GeV}\leq \Delta_{s,u}\leq 0.6\,\mbox{GeV}$ in the parametrization corresponding to the case of an off-shell $B^{\ast}$ and $K$. As can be seen in Fig.~\ref{Figure6}, this procedure gives us an uncertainty interval of $ 20.55\,\mbox{GeV} \leq g^{B^{\ast}}_{B^{\ast}_{s1}B^{\ast}K} \leq 29.73\,\mbox{GeV}$ and $ 19.33\,\mbox{GeV} \leq g^{K}_{B^{\ast}_{s1}B^{\ast}K} \leq 27.87\,\mbox{GeV}$. Taking into account uncertainties due to the continuum threshold parameter, we finally obtain
\begin{eqnarray}
g_{B^{\ast}_{s1}B^{\ast}K}=(24.53\pm5.20)~\mbox{GeV}.
\end{eqnarray}

\begin{table}
\caption{ Theoretical estimations of the strong coupling constants
from different models}
\begin{center}
\begin{tabular}{cccccccc}
\hline\hline
& Coupling constant & This work & \cite{Guo1,Guo2} & \cite{Wang2} & \\
\hline
&$g_{B^{\ast}_{s1}B^{\ast}K}(\mbox{GeV})$  & $24.53\pm5.20$& 23.572  & $18.1\pm6.1$  &\\ \hline
\hline
\end{tabular}
\end{center}
\label{table2}
\end{table}

Together with the predictions from other theoretical approaches, the numerical results of the coupling constant are presented in Table \ref{table2}. Based on the heavy-light chiral lagrangian, the unitarized two-meson scattering amplitudes are analyzed, and the strong coupling constants $g_{B^{\ast}_{s1}B^{\ast}K}=23.572\,\mbox{GeV}$ is obtained~\cite{Guo1,Guo2}. Our result is in good agreement with their estimate. Whereas, it is estimated to be $18.1\pm6.1\,\mbox{GeV}$  with light-cone QCD sum rules~\cite{Wang2}, the central value of which is about $30\%$ smaller than above two predictions.

In conclusion, we have used three-point QCD sum rules to calculate the form factor of the $B^{\ast}_{s1}B^{\ast} K$ vertex. Both cases that $B^{\ast}$ is off-shell and $K$ is off-shell have been considered. As a side product of the form factor, the coupling constant is estimated, which is compatible with the results in Refs.~\cite{Guo1,Guo2} but larger than that from light-cone QCD sum rules~\cite{Wang2}.
\begin{acknowledgments}
The author Chun-Yu Cui thank Yi-Sheng Yang for his useful discussions. This work was supported in part by the National Natural Science
Foundation of China under Contract Nos.10975184 and 11047117, 11105222.
\end{acknowledgments}

\end{document}